\documentstyle[osa,12pt]{revtex}

\begin{document}

\begin{titlepage}

\title{Extracting Work From A Single Heat Bath}

\author{Arbab A. Khan}
\address{\normalsize Department of Electronics, Quaid-i-Azam University, Islamabad, Pakistan}
\author{M.S. Ansari }
\address{\normalsize Department of Chemistry, Quaid-i-Azam University, Islamabad, Pakistan}

\maketitle

\vspace{1in}
\begin{abstract}
We present here a machine that is capable of extracting work from a single heat bath. 
Although no significant temperature gradient is involved in the operation of the 
machine, yet the Carnot efficiency as high as one is achievable. Working of the 
machine is explained on the basis of a demon suggested by Maxwell. Utilizing the 
kinetic energy spectrum of the molecules in solution, the demon can send "hotter" 
molecules to a higher gravitational potential at the expense of their own energies. 
Difference in chemical potentials due to concentration gradients and use of 
semi-permeable membranes ensure the continuing cyclic process.
\end{abstract}

\end{titlepage}


Perpetual Motion Machine ($PMM$)\cite{Angrist,Kauzmann}, although hardly
allowed by the laws of thermodynamics, has been a subject of great interest
since long. $PMM$ of first kind, in contrast to the law of conservation of
energy, is supposed to create energy, thus violating the first law of
thermodynamics. On the other hand $PMM$ of second kind should be capable to
convert total heat into useful work showing 100\% efficiency of the heat
engine, which is rather contrary to the second law of thermodynamics.

Another manifestation that forbids $PMM$ of second kind is that no work can
be extracted from a single heat bath. This is because according to Carnot,
greater efficiency of a heat engine is only possible if temperature of the
energy source, $T_{h}$ is higher and the entropy sink is maintained at a
lower temperature $T_{c}$\cite{Kauzmann,Atkins,Annamalai} whereby the Carnot
efficiency, is given as%
\begin{equation}
\eta =1-T_{c}/T_{h}.
\end{equation}

It is to note further that input of heat energy from a bath is essential for
all heat engines but for a $PMM$ of second kind it can take place without
requiring any temperature difference and entropy sink.

An additional requirement that prohibits the existence of $PMM$ is
pertaining to entropy, which for a spontaneous process must not decrease.
Since $PMM$ of second kind is capable of increasing order by separating high
and low energy particles without any energy involvement, its existence is
not allowed according to the entropy considerations as well. Maxwell's demon %
\cite{Leff,Baeyer}, a hypothetical ``creature'', can sort hot and cold types
of particles without any effort in measurement and consumption of energy,
therefore it may also be considered equivalent to a $PMM$ of second kind.

Discovery of quantum non-demolition measurements and computing\cite%
{Brune,Haroche,Arbab,Imoto,Roch,Zubairy} have given an incentive and led to
a refined review of thermodynamics demanding a more careful search for the
existence of $PMM$.

Recently in their excellent paper Scully et al.\cite{Agarwal} working on a
new kind of quantum heat engine using ``phaseonium''\cite{Scully} as fuel
and utilizing lasing without inversion\cite{Kocharovskaya,Arimondo,Harris},
showed that work can be extracted from a single heat bath but they had to
consider vanishing quantum coherence to save laws of thermodynamics. In a
recent communication Scully\cite{M. Scully} used a Stern-Gerlach apparatus
to sort hot and cold spin atoms thus acting as Maxwell's demon. To achieve a
cyclic operation, the atomic center of mass was prepared in a well-defined
quantum state, showing that the ``cost'' of preparing the center of mass
wave packet was enough to preserve the second law of thermodynamics.

Distribution of kinetic energy between the particles in a sample of gas or
liquid at a given temperature is Gaussian\cite{Reif,Lachish}. In an isolated
system, one possible way of sorting high and low energy particles can be
moving up high-energy particles to a higher gravitational potential at the
expense of their own heat energies. The above scheme may be employed for a
cyclic process of mass-transfer to achieve useful work without involvement
of an initial temperature gradient.

At this point a system of two containers at different gravitational
potentials may be considered as shown in Fig. 1. The container $L$ at lower
potential is filled with neat water, while the container $H$ at higher
potential is initially empty. The system is in thermal equilibrium at an
absolute temperature $T$. The demon allows transfer of higher kinetic energy
particles from $L$ to $H$ utilizing their own energies. Consequently, mass $%
m $ of water may reach to $H$ after gaining the gravitational potential
energy $mgh$, where $g$ is the (average) gravitational constant and $h$ is
the height to which water is raised from the surface of water in $L$. This
will result in an overall decrease in temperature of the thermodynamically
isolated system. This temperature decrease may be related with potential
energy as 
\begin{equation}
\Delta T=mgh/M\sigma ,
\end{equation}%
where $M$ and $\sigma $ are the mass and specific heat capacity of the whole
system. The temperature of the system may again be raised back to $T$ if
water from $H$ is allowed to drain into container $L$. If the setup is
connected to a heat reservoir that can maintain temperature $T$, this demon
will continue to pour more and more water into $H$ by increasing the
potential energy of water at the expense of heat taken from the bath. 
\begin{figure}[tbp]
\caption{Maxwell's demon capable of separating higher energy water molecules
to a container at higher potential.}
\end{figure}

\bigskip Kinetic energy spectrum of molecules in a solution requires the
lighter solvent molecules to be more agile and hence an effective means of
energy transfer. A semi-permeable membrane ($SPM$) allows only solvent
molecules to pass through it but solute molecules, especially of larger
size, are not allowed to cross. Statistically the solvent molecules may move
across both sides of $SPM$. However, a net flux of solvent across the $SPM$
may take place under the influence of gravitational and/or osmotic pressure.

Consider the setup for osmosis as shown in Fig. 2. (A). A tube whose lower
end is attached to an $SPM$ contains sugar solution in it. If the $SPM$ is
immersed in neat water and the system is left alone to equilibrate, the
level of the solution rises in the tube due to osmosis till a height $h$ is
finally attained at which the column level in the tube is maintained by a
counterbalance of water-flow across the $SPM$. Under this condition the
chemical potential of water on the two sides of the $SPM$ becomes equal\cite%
{Tabor}. The column pressure, $\rho gh$ ($\rho $ being the solution density
in appropriate units) makes water to flow out of the tube through $SPM$
while the osmotic pressure tends to maintain the level by an inwards flow.
As a result, a state of dynamic equilibrium is reached. The osmotic
pressure, $\Pi $ may be related to the mole fraction of water, $X_{A}$ as

\begin{equation}
\Pi =-(RT/V_{m})lnX_{A},
\end{equation}

where $R$ is the gas constant, $T$ is absolute temperature and $V_{m}$ is
the molar volume of water. For dilute solutions, eq. (3) simplifies to
well-known van't Hoff's expression that directly relates osmotic pressure
with the solute concentration \cite{Tabor, W. Atkins, Hoff}. It is evident
that the column pressure, $\rho gh$ alone is responsible for the outwards
flow of water through $SPM$. Hence in the event where the $SPM$ is not
immersed, water will come out of it under the influence of the column
pressure. This is depicted in Fig. 2 (B), where shape of the $SPM$ is
slightly modified as per requirement of the experiment to be discussed later.

\begin{figure}[tbph]
\caption{(A) Setup for osmosis showing a rise of the solution level due to
inwards flow of water through the SPM. (B) Setup for the outwards flow of
water under the influence of column pressure, $\protect\rho gl$.}
\end{figure}

Now we join together the above two setups so that the sugar solution becomes
enclosed in a tube whose both ends are attached to $SPM$ as shown in Fig. 3.
Since the lower $SPM$ remains immersed in water in $L$, osmosis continues
and tends to maintain the column height $h$. The inward pressure at the
lower $SPM$ is still governed by eq. (3). On the other hand, the upper $SPM$
has been kept hanging at an average height $k$ above the water level in $L$.
An outwards pressure on the upper $SPM$ is an essential requirement for the
outflow of water through it; keeping $k<h$ ensures this working condition. A
suitable arrangement for the escape of dissolved air or the air that enters
into the tube through $SPM$ may be made by providing an ``air escape valve''
above height $h$ as shown in Fig. 3. Water uptake through the lower $SPM$
and release through the upper $SPM$ makes reaching a dynamic equilibrium at
which a steady flow of water from $L$ to $H$ is maintained. The neat water
collected in $H$ may be made to drain back into $L$ thus completing the flow
cycle and providing an opportunity to extract useful work from the system.
We have successfully achieved the working of the setup shown in Fig. 3 at
ambient temperature and also repeated the experiment. For this purpose,
about 30 \% by weight sugar solution was used and a natural $SPM$ was
employed. 
\begin{figure}[tbp]
\caption{Setup showing a PMM of second kind, regarded as a Maxwell's demon,
which can totally convert absorbed heat into useful work without any
involvement of temperature difference.}
\end{figure}

Functioning of the above $PMM$ depends upon various factors related to its
design and the working conditions. A steady flow rate, besides other
factors, depends upon quality \& size of the $SPM$, concentration of the
sugar solution and temperature of the system. A number of factors tend to
hinder the performance of the $PMM$. One important factor is the
accumulation of water droplets at the outer surface of the upper $SPM$. This
problem may be minimized by properly designing the shape of the $SPM$.
Another factor is the concentration gradient developed within the solution
due to induction of neat water. Diffusion of sugar molecules and a slow rate
of flow tend to nullify this gradient. The net effect of both the above
factors is a reduction of the flow rate than expected.

We conclude that the setup shown in Fig. 3 behaves as a Maxwell's demon
without itself consuming any energy or involving any change of state, but
converts heat into potential energy. It is notable that the whole system is
at a single temperature, $T$ maintained by the heat bath of the surrounding.
Further the system is non-temporal, i.e., all the parameters like
concentration and pressure etc. are time independent.

\bigskip {\bf Figure Caption%
\vspace{0.5in}%
}

Fig. $1$ Maxwell's demon capable of separating higher energy water molecules
to a container at higher potential.

Fig. $2$ (A) Setup for osmosis showing a rise of the solution level due to
inwards flow of water through the SPM. (B) Setup for the outwards flow of
water under the influence of column pressure, $\rho gl$.

Fig. $3$ Setup showing a PMM of second kind, regarded as a Maxwell's demon,
which can totally convert absorbed heat into useful work without any
involvement of temperature difference.

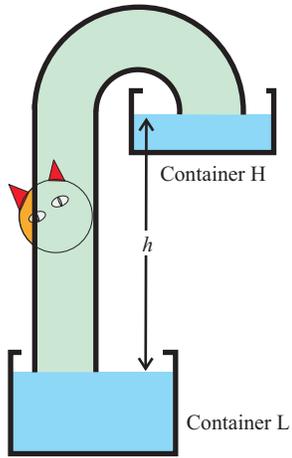
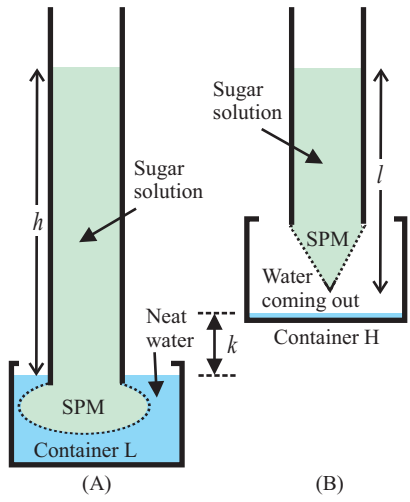
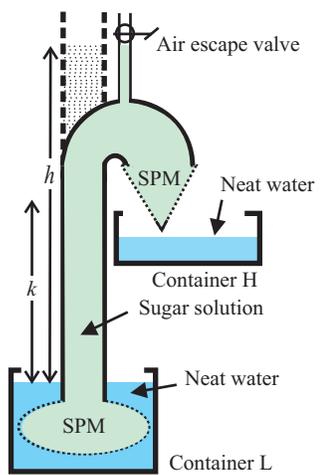
\end{document}